**Title**: The Effects of Higher Education on Midlife Depression: Quasi-Experimental Evidence from South Korea

**Authors (Name/Institutions/Country or regions/Email address)**

Ah-Reum Lee*, Department of Epidemiology and Biostatistics, University of California, San Francisco, USA (ahreum.lee@ucsf.edu)

Jacqueline M. Torres, Department of Epidemiology and Biostatistics, University of California, San Francisco, USA (jacqueline.torres@ucsf.edu)

Jinkook Lee, Research Professor of Economics, Director of the Program on Global Aging, Health, and Policy, University of Southern California (jinkookl@usc.edu)

*Corresponding author
Address: 505 Parnassus Ave, San Francisco, CA 94143, United States

**Abstract**

Higher education has expanded worldwide, with women outpacing men in many regions. While educational attainment is consistently linked to better physical health, its mental health effects – particularly for women – remain underexplored, and causal evidence is limited.

We estimate the impact of college completion on depression among middle-aged women in South Korea, leveraging the 1993 higher-education reform, which raised women's college attainment by 45 percentage points (pp) over the following decade. We use two nationally representative datasets to triangulate evidence: the Korea National Health and Nutrition Examination Survey (KNHANES, 2007-2021) for physician-diagnosed depression, and the Korean Longitudinal Survey of Women and Families (KLoWF, 2007-2022) to validate findings using self-reports of depressive symptoms. We implement two-stage least squares (2SLS) with a birth-cohort instrument based on exposure to the reform (±3-year cohorts in KNHANES; ±1–3-year cohorts in KLoWF).

In KNHANES, college completion lowers physician-diagnosed depression by 2.4 pp, attenuating to 1.6 pp after adjusting for income, employment, and physical health. In KLoWF, college completion improves self-reported mental health: the weekly depressive-symptoms composite declines by 17.4 pp, attenuating to 16.4 pp after covariate adjustment. Placebo tests on unaffected cohorts yield null results.

This study contributes to the growing quasi-experimental literature on education and mental health with convergent evidence across clinical diagnoses and self-reported depressive symptoms in South Korea. By focusing on college education in a non-Western setting, it extends the external validity of existing findings and highlights educational policy as a potential lever to reduce the burden of midlife depression among women.

## Highlights

- We use the 1993 Korean higher-education reform as an IV for college effects.

- College lowers both diagnosed depression and self-reported depressive symptoms.

- Economic and physical health partly mediate; behaviors contribute little.

- Results support tertiary education as a structural lever for mental health equity.

# 1. Introduction

Over the past decades, global access to tertiary education has expanded significantly, with nearly 40% of the world's population now having access to higher education (1). Women, in particular, have benefited from this expansion, often outpacing men in college enrollment and degree attainment in many regions (2). While the benefits of education for economic and social mobility are well established (3,4), its implications for health – especially mental health for women – are less understood. Higher educational attainment has been associated with lower mortality (5-16), lower rates of chronic conditions such as diabetes and cardiovascular disease (17,18), better physical functioning (19,20), and healthier behaviors (17,21). More recently, attention has shifted to the potential mental health benefits of education, with studies suggesting that individuals with more years of schooling tend to report lower rates of depression and anxiety, greater emotional resilience, and stronger psychological well-being (17,22,23). Yet despite this consistent observational evidence, the mechanisms linking higher education and mental health are not well understood, and relatively little causal research has been conducted in this area– particularly in quasi-experimental frameworks.

Education can shape mental health outcomes through multiple reinforcing pathways, providing individuals with the resources, autonomy, and resilience necessary to maintain overall well-being (17). Higher educational attainment – especially completion of college or other forms of tertiary education – reduces exposure to economic stressors, unstable employment, and adverse environments (24), all of which are known risk factors for psychological distress and depression (25). While basic education offers foundational skills such as literacy and numeracy that may help individuals meet everyday demands, higher education further equips them with the cognitive tools to navigate complex systems, including healthcare. This, in turn, improves access to preventive services and treatment and fosters more informed, proactive decision-making about personal health (26). Beyond healthcare access, education enhances psychological resilience by promoting a sense of control and self-efficacy, both of which are associated with lower stress and a reduced risk of depression (22). More highly educated individuals are also more likely to have strong social networks that offer emotional and financial support, buffering against life stressors and mitigating mental distress (27). Furthermore, they tend to exhibit more adaptive responses to adversity, reducing the long-term psychological toll of negative life events (28,29). These psychological and social resources position education as a potentially powerful upstream determinant of mental health.

However, the causal nature of the relationship between education and mental health remains debated (30-33). First, reverse causality is a key concern – early-life mental health challenges can disrupt academic performance and lower educational attainment. For example, adolescents with depressive symptoms are less likely to complete high school or pursue postsecondary education, reinforcing socioeconomic disadvantage across the life course (34,35). Second, confounding factors such as family background, early-life adversity, and neighborhood environment may influence both education and mental health, making it difficult to disentangle the effects of schooling from broader social determinants. Moreover, even where education is associated with better mental health, some evidence suggests that its protective effects may weaken over time or follows a curvilinear pattern with larger effects among young adults and the elderly (36,37).

To address these concerns, researchers have increasingly used exogenous variation in schooling from compulsory schooling laws to study the relationship between education and health. Evidence from these studies suggests that higher educational attainment has a positive causal impact on physical health, contributing to lower mortality (7,9,30,37-47) and better physical health (42,49). These benefits are often attributed to mechanisms such as greater economic opportunities, healthier lifestyles, and improved working conditions (42,43). However, the magnitude of these effects appears to vary by gender, with some studies suggesting smaller or even negligible health benefits for women, potentially due to offsetting factors such as marital dynamics and caregiving responsibilities (42,43).

Fewer quasi-experimental studies have examined its impact on mental health. One exception is a recent study from Zimbabwe (48), which leveraged age-specific exposure to an education reform as an IV and found that increased schooling significantly reduced depressive symptoms, particularly among women and rural residents. However, beyond this example, there remains limited causal evidence on how education affects mental health outcomes in other global contexts.

Furthermore, most studies have focused on the health effects of primary or lower-secondary school completion, leaving the effects of college education – particularly for women who have recently gained broader access to higher education – largely underexplored. Most existing evidence on higher education comes from the U.S. (36, 41-46), Canada (47), and Germany (49), and relies on instruments such as draft risk during the Vietnam War (41) or regional variation in college accessibility (42,44,45), as well as other quasi-experimental approaches, such as twin comparisons (44) and microsimulations (47). However, the external validity of this literature remains limited, and the mental health dimension has received comparatively little attention.

In this study, we investigate the impact of higher education on mental health and its underlying mechanisms among middle-aged women in South Korea, with a particular focus on depressive symptoms. Depression is a growing public health concern in South Korea, especially among women, who exhibit higher prevalence than men across most age groups (50). South Korea offers a compelling setting to examine the causal effects of higher education on women's mental health due to its dramatic expansion in access to higher education in recent decades. A pivotal moment in this expansion was the 1993 higher education reform, which abolished strict admission quotas and led to a surge in female college enrollment (51). This policy shift has been used to estimate the causal effects of higher education on women's fertility (52), labor force participation, and earnings (53). We build on this work by leveraging the same reform as a quasi-experimental source of exogenous variation in educational attainment to estimate the causal impact of college education on mental health.

Our analysis draws on two complementary, nationally representative datasets: the Korea National Health and Nutrition Examination Survey (KNHANES, 2007–2021) and the Korean Longitudinal Survey of Women and Families (KLoWF, 2007–2022). KNHANES provides indicators for physician-diagnosed depression alongside rich covariates on health behaviors and socioeconomic status. KLoWF, in contrast, offers self-reports of depressive symptoms and detailed measures of daily affect and stress. Leveraging cohort variation created by the reform – where those born after the threshold faced markedly expanded college access relative to adjacent cohorts – we estimate the effect of education on depression among women using a two-stage least squares instrumental variable (2SLS/IV) design (54). In KNHANES, we define exposure with broader ±3-year birth-cohort bands around the 1974 cutoff. In KLoWF, the larger panel affords substantially narrower bands (±1-3 years), which sharpen the instrument by focusing comparisons on cohorts that are very similar in both individual characteristics and secular conditions, thereby reducing cohort-related confounding.

Informed by Fundamental Cause Theory (56,57), we hypothesize that college attendance reduces the risk of depression by increasing access to "*flexible resources*" – such as stable employment, higher income, and health-promoting behaviors and knowledge – that protect mental health. To test this hypothesis, we examine the mediating roles of economic conditions, health behaviors, and physical health status in the relationship between education and depression. We present estimates both before and after adjusting for these potential mediators and further examine the mediators as secondary outcomes within a two-stage least squares (2SLS) framework to clarify their role in the broader education-health relationship.

We find that completing college reduces the likelihood of being diagnosed with depression by 2.4 percentage points (pp), attenuating to 1.6 pp after adjusting for potential mediators (KNHANES). This estimate may be conservative, as it is based on physician diagnoses, which may not capture subclinical or undiagnosed cases – particularly among lower-income women with limited healthcare access. The effect is partially explained by improvements in income, employment, and physical health, while health behaviors such as smoking, drinking, and preventive care play a smaller role. A sizable portion of the effect remains unexplained, suggesting additional pathways. Robustness checks, including placebo analyses on unaffected cohorts and controls for cohort-specific trends, yield null effects, reinforcing the validity of our results.

We confirm the pattern in KLoWF using self-reported depressive-symptom indices with tighter cohort windows (±1 year). IV estimates indicate that the weekly depressive-symptoms composite falls by about 16 pp. Broadening the cohort window to ±2/±3 years generally yields larger reductions in depressive symptoms (±2 year cohorts: ~19 pp vs ±3 year cohorts: ~30 pp), with an improvement in daily affect/stress. Consistent with KNHANES, economic conditions (- 1.2 pp) and health behaviors (-0.9 pp) partially mediated the effect, which remained largely unchanged. Because these mental-health outcomes are subjective, self-reported measures, they may reflect both true symptom changes and reporting or perception shifts; nevertheless, patterns are consistent across multiple measures of depressive symptoms and diagnoses and across varying cohort bounds.

However, several limitations warrant caution. First, our sample includes only middle-aged South Korean women, which may limit generalizability. Second, although we employ a quasi-experimental design, unobserved confounders – especially events concurrent with the reform – may bias estimates. Lastly, we were also unable to test subgroup effects (e.g., urban vs. rural) due to power and measurement constraints. Nevertheless, our results point to meaningful mental health returns to college education. Factoring these into cost-benefit calculations strengthens the rationale for policies that expand access to higher education, particularly for women.

The remainder of the paper is organized as follows. Section 2 provides institutional background on the 1993 higher education reform in South Korea. Section 3 describes the data sources and empirical strategy. Section 4 presents the main results, followed by additional analyses and robustness checks. Section 5 offers a discussion of the findings and their limitations. The final section concludes.

## 2. Institutional Background

Our analytic strategy exploits variation in the college education induced by the 1993 higher education reform in South Korea, initiated by President Kim Young-Sam, which marked a strategic shift from a quota-based, government-regulated system to one emphasizing deregulation and institutional autonomy (58,59). At the heart of this transformation were two key policies: the "Plan for the Promotion of College Enrollment Liberalization" and the "5·31 Education Reform Plan" (60). These policies expanded university enrollment capacity, allowing new institutions to operate with relaxed admission standards. As a result, between 1993 and 1997, college enrollment rates surged by over 25 percentage points (61). Furthermore, the reform encouraged the growth of specialized colleges, reflecting the government's commitment to diversifying educational pathways (61).

One of the most striking outcomes of these reforms was the dramatic rise in female college enrollment. The percentage of women attending higher education in South Korea increased from 22% in 1980 to 33% in 1990, and following the 1990s reforms, surged to 80% by 2004 – more than doubling within a decade (62). Our identification strategy relies on this variation in college attendance and completion over time to isolate the causal effect of education on women's health outcomes.

## 3. Data and Methods

### 3.1. Data

#### 3.1.1. Korea National Health and Nutrition Examination Survey (KNHANES)

**Survey Overview**. This study uses data from the Korea National Health and Nutrition Examination Survey (KNHANES), a national system assessing Koreans' health and nutrition since 1998 (55). KNHANES is a nationally representative cross-sectional survey that samples about 10,000 individuals annually. It collects data on socioeconomic status, health behaviors, quality of life, healthcare utilization, anthropometric measures, clinical profiles, and dietary intakes through health interviews, examinations, and nutrition surveys. Initially conducted in 1998, 2001, and 2005, the survey transitioned to an annual program from 2007 onwards.

**Sample Construction**. Our analytic sample is drawn from the 2009-2021 KNHANES, a nationally representative, pooled cross-sectional dataset. We begin in 2009 because several key variables, such as parental education, depression diagnosis, and health behaviors, were either unavailable or inconsistently measured in earlier waves. We restricted the sample to women aged 40 to 65 (n = 28,236), as this group aligns with our study's focus on midlife mental health. We further narrowed the sample to include only women born within five years before or after the 1993 Higher Education Reform (n = 3,211), which is how we define our instrumental variable. After excluding respondents with missing data on relevant covariates, exposures, and outcomes, the final analytic sample was 2,912 women (n = 1,132 treatment group; n = 1,780 control group). In the sensitivity analyses, we vary birth cohort bounds (3-year, 5-year, 10-year, and 15-year) to assess how different proximity to the reform impacts outcomes.

**Exposure.** Our primary exposure variable measures whether an individual has attained higher education, specifically completing a 2-year/3-year college or a 4-year university. Respondents reported their highest level of education based on the following categories: traditional village school/classical Chinese studies, no formal education, elementary school, middle school, high school, 2-year/3-year college, 4-year university, graduate school, not applicable (e.g., young children), and unknown/no response. The variable is coded as 1 if the respondent has completed a 2-year/3-year college, a 4-year university, or higher, and 0 otherwise. Those who reported missing or inapplicable responses were excluded from the analytic sample (n=251).

**Primary Outcomes.** Our primary outcome is whether the respondent has ever been diagnosed with depression, based on a binary indicator of self-reported physician diagnosis. The variable is coded as 1 if the respondent reports a diagnosis and 0 if not (original response options: 0 = No, 1 = Yes, 8 = Not applicable [adolescent/child], 9 = Don't know/No response). While this measure offers a clear and clinically relevant indicator of depression that is less prone to subjective interpretation than symptom-based self-assessments, it may nonetheless underestimate the true prevalence and severity of mental health conditions, particularly among individuals who remain undiagnosed due to limited access to physicians or psychiatrists and stigma surrounding mental healthcare (64). **Figure 1** shows the distribution of this outcome in the study population.

**Secondary Outcomes**. We examine three sets of secondary outcomes that may serve as potential mediators in the relationship between higher education and mental health: (1) economic conditions, including household income (inflation-adjusted, in units of 10,000 KRW) and full-time employment status; (2) health behaviors, such as current smoking, current drinking, and use of preventive services (e.g., health check-ups in the past two years); and (3) physical health, including self-reported health (average or above vs poor or below), body mass index (BMI), and number of pregnancies. See **Appendix 1** for full variable definitions and coding.

**Covariates**. Our analysis includes a range of demographic, socioeconomic, and family-level characteristics. Specifically, we control for key sociodemographic variables, including age (in years), marital status (married or partnered vs. separated, divorced, widowed, or never married), and occupational category (as described above). Family-level variables include household size, household income, and region of residence (covering 17 provinces and special cities). To capture early-life circumstances that may influence both educational attainment and later health, we include father's education as a proxy for socioeconomic background – a well-established determinant of early childhood development and educational outcomes (66), and an indicator for family history of chronic conditions (hypertension, hyperlipidemia, ischemic heart disease [myocardial infarction or angina], stroke, diabetes, thyroid disease, or hepatitis B). Lastly, we adjust for potential mediators described in the secondary outcomes section to help isolate the direct effects of education from health-related lifestyle factors.

**Instrumental Variable.** The reform substantially increased access to higher education, particularly for individuals born in 1974 or later (**Figure 2**). Given that exposure to expanded college opportunities varied across birth cohorts, birth year serves as a plausibly exogenous determinant of educational attainment. We leverage birth cohort membership as an instrument to identify individuals whose educational decisions were influenced by the expansion. To operationalize this, we define our instrumental variable (IV) as follows: the treatment group consists of individuals born between 1974 and 1976, who were directly impacted by the policy expansion and are assigned a value of one. The control group includes those born before 1973, within the preceding three-year period, who were unaffected by the reform and are assigned a value of zero.

The validity of our IV approach relies on several key assumptions. First, the reform must have significantly influenced access to higher education, ensuring the relevance assumption holds (54). The 1993 policy expansion altered the likelihood of college attendance based on birth cohort, generating clear variation in educational attainment (**Figure 2**). Another critical assumption is monotonicity (54,63), which implies that no individual experienced a reduction in their likelihood of attending college due to the reform. Under these conditions, the IV estimate isolates the effect of education on health for compliers (i.e., individuals whose college attendance was induced by the reform) (54,63).

Our identification strategy assumes that a substantial portion of individuals who pursued higher education in the post-reform period did so as a direct result of the policy change, classifying them as compliers. South Korea's higher education expansion provides a particularly strong setting for this approach. Unlike many other contexts where policy-driven increases in college access were gradual, South Korea's reform led to a dramatic and rapid transformation, shifting college graduates from a minority to a majority within a short period (53). This large-scale change strengthens the instrument's exogeneity and relevance, ensuring that the observed variation in education is driven by policy rather than unobserved confounders. To establish the relevance of our instrument, we provide formal econometric evidence in **Table 1** (Partial F=15), where we assess the first-stage relationship between birth cohort membership and educational attainment. Prior studies also confirm that the 1993 reform significantly expanded access to higher education for cohorts born after 1974, as shown by analyses of census data (52).

Additionally, the reform should affect later-life health only through its impact on education rather than via alternative channels, such as changes in the quality of higher education or broader secular and macroeconomic trends. Several considerations support this exclusion restriction. First, although rapid expansion of higher education could have degraded institutional quality, prior evidence indicates that newly established colleges were comparable to existing institutions on key inputs (e.g., student-faculty ratios, library resources, per-student investment), making quality-related pathways to mental health unlikely (53). Second, concerns about secular trends, including macroeconomic dynamics, are limited: South Korea's growth was relatively stable through much of the 1990s, and although the 1997 Asian Financial Crisis introduced substantial disruption, there is no clear evidence that these shocks differentially affected cohorts just above versus below the reform cutoff in a way that would bias our estimates. To disentangle the effects of education from secular trends, we conduct several robustness checks, using placebo groups, varying birth-cohort windows, and cohort-trend interactions (see **Methods** for details on these robustness checks). We also replicate our main findings in an independent dataset (KLoWF) using self-reported symptom indices and an instrument based on narrower birth-cohort bands (±1–3 years).

### 3.1.2. Korean Longitudinal Survey of Women and Families (KLoWF)

**Survey Overview**. KLoWF is a nationally representative longitudinal panel of women in Korea, fielded biennially since 2007, designed to track socio-demographic conditions, employment, health and well-being, family structure, and psychosocial status (67). Core modules include detailed mental-health items (weekly symptoms, daily affect and stressors, suicidal ideation/attempts, and stress-coping), health-related quality of life, and early-life background. We analyze waves 4–9 (2012–2022), beginning in 2012 when the survey introduced the detailed mental-health battery (weekly depressive-symptom items and daily affect/stress indices).

**Sample Construction**. We mirror the KNHANES cohort window, restricting to women born within ±1 year of the 1974 cutoff – the first cohort exposed to the 1993 Higher Education Reform – and excluding observations with missing exposure, outcomes, or listed covariates (n=22); the final analytic sample includes 751 women contributing 4,396 person-wave observations. As sensitivity checks, we replicate the analyses using ±2-year and ±3-year windows, yielding 1,554 individuals (5,674 person-wave observations) and 2,317 individuals (8,314 person-wave observations,) respectively.

**Exposure**. Education is harmonized with the KNHANES. We construct a binary indicator for college completion equal to 1 if a respondent reports completing a 2–3-year college, a 4-year university, or graduate school; 0 if the highest attainment is at most high school.

**Primary Outcomes**. We measure depressive symptoms using complementary indicators with different recall windows. First, we construct a weekly symptom composite by averaging responses to 10 items referring to the prior week: feeling bothered, difficulty concentrating, feeling very depressed, finding everything very hard, fearfulness, sleep disturbance, loneliness, low energy, and two positively worded items ("did okay" and "no major complaints"), which closely align with CES-D short-form content (68). Items are originally recorded on a 4-point Likert scale (1=strongly agree to 4=strongly disagree), and we reverse-code the negatively worded items so that higher values indicate worse depressive symptoms. For interpretability, we create binary indicators for each depressive-symptom item, coded 1 if the respondent reports strongly agree/agree and 0 otherwise; the binary composite is the mean of these indicators.

Additionally, we construct a daily affect/stress index from the "daily thoughts and emotions" module, which asks about eight experiences: feeling stressed, drinking alone because of stress, preferring solitude, becoming angry, having no one to talk to, managing stress well, experiencing financial stress, and experiencing interpersonal stress. Items are also recorded on a 4-point Likert scale and are reverse-coded so that higher values uniformly indicate worse affective states (excluding the positively framed "managing stress well" item). For interpretability, we again create binary indicators for each item (1 = endorses the adverse state; 0 otherwise) and a binary composite equal to the mean of these item-level binaries.

**Secondary Outcomes**. We construct secondary outcomes that align with the domains analyzed in the KNHANES. Specifically, we examine three sets of potential mediators linking higher education to mental health: (1) economic conditions, including household income and full-time employment status; (2) health behaviors, including smoking, receipt of preventive health care (e.g., recent health check-up), and frequent vigorous exercise (three or more days per week); and

(3) physical health, including self-rated health, number of pregnancies, and number of chronic conditions. See **Appendix 1** for full variable definitions and coding. Relative to the KNHANES, the KLoWF does not include alcohol use in a directly comparable form, but it does capture frequent exercise; similarly, it does not include body mass index, but it does measure chronic disease burden.

**Instrumental Variable**. We exploit the same higher-education reform but take advantage of the panel's larger sample to tighten cohort windows. Our main specification instruments college completion with indicators for being born in the post-reform cohorts using ±1–3-year windows (vs. 3-year windows for KNHANES); the main analysis uses the narrowest window that preserves adequate first-stage strength (i.e., ±1 year) and report ±2- and ±3- year windows as sensitivity analyses.

**Covariates**. We adjust for several early-life and family background characteristics that may influence both educational attainment and adult mental health, including number of siblings; birthplace urbanicity (large metropolitan area, mid-sized or small city, rural area, or overseas); mother's education around age 15 (high school graduate or higher vs. less); family financial situation around age 15 (above average vs. poor or below); and parental relationship quality (above average vs. poor or below). Additional details are available in **Appendix 1**.

## 3.2 Empirical Strategy

**Statistical Analysis**. To investigate the effects of respondents' college completion on their mental health, we employ a conventional two-stage least squares (2SLS) estimation with an instrumental variable (IV) approach. This method helps address potential endogeneity by leveraging variation in educational attainment induced by the 1993 higher education reform in South Korea. In the first stage, we estimate the likelihood of college completion as a function of exposure to the reform and a set of covariates:

$$1^{st} \text{ stage: } College\ Completion_i = \gamma + \varphi_1 1993_{reform_i} + \varphi_2 Control_i + v_i$$

where $College\ Completion_i$ represents whether respondent $i$ completed 2-year/3-year college or a 4-year university (coded as 1 if completed, 0 otherwise), and $1993_{reform}$ is a binary indicator equal to 1 if the respondent belongs to cohorts exposed to the reform – birth years within three years after 1974 in KNHANES (1974–1976) and within one year in KLoWF (1974) – and 0 for adjacent pre-reform cohorts that just missed exposure – three years before 1974 in KNHANES (1971–1973) and one year before 1974 in KLoWF (1973); $Control_i$ is a vector of individual and family level covariates, as detailed in the Data section above; and $v_i$ is the error term. $\varphi_1$ is the primary coefficient of interest, which captures the increase in college completion rates associated with the 1993 higher education reform.

In the second stage, we model the relationship between educational attainment and mental health outcomes, using the predicted values of college completion from the first stage as an instrument:

$$2^{nd} \text{ stage: } Depression_{it} = \alpha + \beta_1 \widehat{College\ Completion}_i + \beta_2 Control_{it} + \varepsilon_{it}$$

where $Depression_{it}$ is a mental health outcome for individual $i$ at time $t$. In KNHANES, $Depression_{it}$ is a binary indicator of whether the respondent has ever been diagnosed with depression by a physician; in KLoWF, it is based on self-reported depressive symptoms. The second-stage model includes the same covariates as the first stage.

**Robustness Tests**. A key concern in our analysis is the potential influence of unobserved confounders that may drive the observed relationship between higher education and later-life health. One potential confounder is the broader macroeconomic environment, which varied across birth cohorts and may have independently influenced both educational attainment and long-term health outcomes, separate from the effects of the 1993 higher education reform. This concern is more salient in KNHANES because we use broader ±3-year cohorts.

To explore this possibility, we first conduct a placebo test using cohorts that were unaffected by the reform. Specifically, we compare individuals born between 1964–1968 with those born between 1969–1973 (the original control group). Since neither cohort was exposed to the policy change, this test allows us to assess whether cohort-specific differences in health outcomes – unrelated to education – might bias our estimates. If substantial differences are detected between these groups, it would suggest the presence of unobserved temporal trends that could confound our main findings. We also attempted to compare the control group (1969–1973) to a younger placebo cohort (1984–1988), who were likely unaffected by the reform due to delayed exposure. However, this comparison was infeasible due to limited sample size. For KLoWF, where identification relies on much narrower cohorts, we implement placebo thresholds by shifting the cutoff to nearby years (1972, 1973, 1975, 1976).

Additionally, we incorporated cohort-trend interactions into our primary models in KNHANES to account for macroeconomic shocks that may have disproportionately affected specific age groups. However, we acknowledge that, while this approach helps control for differential cohort trends, it may not fully capture heterogeneity within cohorts, particularly differences based on socioeconomic background or regional economic conditions.

## 4. Results

### 4.1. KNHANES

**Descriptive Statistics**. **Table 2** compares women in three-year birth cohorts before and after exposure to South Korea's 1993 higher-education reform. The treatment group (n=1,132; 38.9%) is younger on average than the control group (n=1,780; 61.1%) (43.0 vs. 45.0 years, p<0.001). Early-life conditions are balanced: father's education and family history of chronic conditions are similar across groups (p=0.127 and p=0.229). Consistent with the reform's intent, college completion is higher in the treatment group (56.0% vs. 47.5%, p<0.001). Marital status and most health behaviors/outcomes are comparable, although the treated report slightly fewer pregnancies (2.7 vs. 2.8, p<0.001), higher monthly household income (546.1 vs. 514.3, p<0.001), and a slightly larger household size (3.6 vs. 3.5, p=0.003).

**First-Stage Results**. The first-stage results indicate that the 1993 higher education reform in South Korea had a strong effect on college completion rates, leading to an approximately 8 percentage point (pp) increase in college attainment among women. The partial F-statistic of 15 exceeds the conventional threshold of 10, suggesting that the IV (exposure to the reform) is strongly correlated with college completion (**Table 1**).

**Second stage results – primary outcomes.** **Table 3** reports second-stage estimates of the effect of college education on the probability of physician-diagnosed depression. We estimate a baseline model (M1) without mediators and then sequentially add mediator sets. In M1, college education is associated with a 2.4 pp reduction ($\beta = -0.024$, 95% CI: $-0.035$, $-0.012$, p<0.001). Adding economic conditions in M2 attenuates the effect to 1.9 pp ($\beta = -0.019$, 95% CI: $-0.030$, $-0.008$, p<0.001), implying partial mediation via improved economic circumstances. Adjusting for health behaviors in M3 yields an estimate similar to baseline ($\beta = -0.022$, 95% CI: $-0.033$, $-0.011$, p<0.001), indicating weak mediation through behaviors. Incorporating physical health in M4 reduces the effect to 2.0pp ($\beta = -0.020$, 95% CI: $-0.030$, $-0.009$, p<0.001), again consistent with partial mediation. In the fully adjusted model (M5), which includes all mediators, the association declines to 1.6 pp (β = −0.016, 95% CI: −0.020, –0.003, p<0.001) yet remains robust, suggesting education's protective effect is only partly explained by economic, behavioral, and health-related pathways.

**Second stage results – secondary outcomes.** To further understand the mechanisms through which education influences health, we examined the potential mediators – economic conditions, key health behaviors, and physical health indicators – as secondary outcomes (**Table 4**). College education is significantly associated with higher household income (β = 43.015, 95% CI: 27.331 58.700). In contrast, there is no clear association with full-time employment status (β = 0.055; 95% CI: −0.048, 0.158), suggesting that while college-educated women may benefit economically through household income, this does not necessarily reflect sustained labor force participation. This pattern is consistent with the experiences of many middle-aged and older women in South Korea, who often face early labor force exit or career interruptions due to caregiving responsibilities or labor market constraints. Moreover, the observed increase in household income may reflect positive assortative mating, whereby college-educated women are more likely to have highly educated – and potentially higher-earning – spouses.

Next, no significant differences are observed in alcohol consumption or self–rated health. However, college education is significantly associated with lower smoking rates (β = –0.014, 95% CI: –0.023, –0.006). In addition, college graduates report significantly fewer pregnancies (β = -0.160, 95% CI: –0.222, –0.098) and lower BMI (β = –0.355, 95% CI: –0.523, –0.186), both of which may partially explain improvements in mental health. The effect on recent health check-up attendance is null (β = 0.012, 95% CI: –0.008, 0.033), suggesting limited influence on preventive care behaviors.

**Robustness checks.** We first compare individuals born between 1968 and 1970 with those born between 1971 and 1973, as neither cohort was affected by the 1993 higher education reform. Since both groups fall outside the exposure window, any differences in health outcomes would likely reflect unobserved cohort-specific trends rather than differences in educational attainment. As shown in **Table 5**, across all model specifications, we find no statistically significant association between cohort membership and depression diagnosis. The estimated coefficients are small in magnitude and statistically indistinguishable from zero in both the unadjusted model ($\beta = –0.005$, 95% CI: –0.014, 0.004) and the fully adjusted model ($\beta = –0.001$, 95% CI: –0.007, 0.009). These null findings provide reassurance that birth cohort effects are unlikely to introduce substantial bias when using the reform-based instrument, though we cannot fully rule out residual cohort-specific differences.

Second, we conducted a robustness check by incorporating cohort-trend interactions into our fully adjusted model to account for potential macroeconomic shocks or unobserved temporal trends that may have differentially impacted specific birth cohorts, independent of educational attainment. As shown in **Table 6**, our main model (M1) estimates a statistically significant 1.6 percentage point reduction in depression diagnosis associated with college education ($\beta = –0.016$, 95% CI: –0.020, –0.003). When cohort-trend interactions are added in Model 2 (M2), the effect size remains identical ($\beta = –0.016$), with a slightly wider confidence interval (95% CI: –0.027, –0.005). These results suggest that our main findings are robust to potential cohort-specific confounding and that the estimated mental health benefits of college education are unlikely to be the main driver of the estimated mental health benefits of college education.

Finally, the results in **Table 7** assess the robustness of our findings by varying the birth cohort windows used to construct the instrument. The main specification with a 3-year window (Model 1) indicates a reduction in depression incidence (β = −0.016; 95% CI: −0.020, −0.003). A wider 5-year window (Model 2), which improves statistical power and cohort comparability but may increase the potential for confounding, yields a smaller, negative estimate (β = −0.009; 95% CI: −0.012, 0.000) that is directionally consistent with the 3-year result. However, expanding the cohort window to 10 years (Model 3) and 15 years (Model 4) produces smaller, statistically insignificant estimates. This attenuation likely reflects increased heterogeneity in reform exposure and unobserved differences across wider birth cohorts.

**4.2. KLoWF**

**Descriptive Statistics**. **Table 8** compares women born in 1973 (control) with those born in 1974 (treated), using the ±1-year window (n=2,780; 50% in each group). As expected, the treated cohort is younger (mean age 39.8 vs 40.7; p<0.001). Treated respondents have slightly more siblings (3.95 vs 3.86; p=0.040) and are more likely to be born in rural areas (metro: 33% vs 37%; rural: 47% vs 42%; p<0.001). Paternal education at age 15 is similar across groups (high school or more: 20% vs 22%; p=0.106), as are family finances and parental relationship quality at age 15 (p>0.20 and p=0.573, respectively).

Adult socioeconomic markers show modest differences: college completion is higher in the treated group (54% vs 45%; p<0.001), marriage is slightly more common (91% vs 89%; p=0.025), and employment is marginally higher (52% vs 49%; p=0.030). See **eTable 1** and **eTable 2** for descriptive statistics for the 2- and 3-year birth-cohort windows.

**First-Stage Results**. Exposure to the 1993 higher-education reform is strongly associated with college completion (**Table 1**). The reform increases college completion by 11.6 pp (95% CI:

0.080, 0.152). The Partial F-statistic of 40 is well above the conventional threshold of 10, indicating a strong first stage.

**Second stage results – primary outcomes. Table 9** and **Table 10** report 2SLS estimates of the effect of college completion on mental health. Results indicate sizable improvements in self-reported depressive symptoms. The weekly depressive-symptoms composite (range 0–1; average of binary items) decreases by 16.4 pp ($\beta = -0.164$, 95% CI: −0.576, −0.005, $p<0.05$). Item-level estimates point in the same direction: feeling very depressed decreases by 20.0 pp ($\beta = -0.200$, 95% CI: −0.805, 0.008, $p<0.10$) and difficulty concentrating decreases by 20.6 pp ($\beta = -0.206$, 95% CI: −0.805, 0.001, $p<0.05$), with additional declines for "felt bothered," "found things very hard," "felt afraid," "felt lonely," and "had no energy." (**Table 9**)

On daily affect and stress, the composite index (0–1) decreases by 16.3 pp ($\beta = -0.163$, 95% CI: −0.691, −0.012, $p<0.10$), with fewer reports of having no one to talk to and improved stress management (reverse-coded) (**Table 10**).

Broadening the cohort window increases precision and yields somewhat larger absolute effects (**eTable 3**), though wider windows may also admit additional confounding from secular trends and cohort composition (potentially weakening the exclusion restriction). With a 2-year window, the depressive-symptoms composite decreases by 19.4 pp ($\beta = -0.194$, 95% CI: −0.297, −0.091, $p<0.001$); in the larger 3-year specification (N = 8,550), the decrease is 31.0 pp ($\beta = -0.310$, 95% CI: −0.434, −0.187, $p<0.001$). For daily affect/stress, the 2-year estimate is close to zero and imprecise ($\beta = -0.089$, 95% CI: −0.210, 0.033), whereas the 3-year window shows a statistically significant improvement of 14.3 pp ($\beta = -0.143$, 95% CI: −0.270, −0.017, $p<0.05$) (**eTable 4**). We interpret these estimates using broader windows as supportive robustness evidence for our main findings, while noting their greater susceptibility to residual confounding.

**Second stage results – secondary outcomes.** To understand potential pathways from education to health, we estimate 2SLS effects on economic conditions, health behaviors, and physical health (**Table 4**). College completion is linked to clear economic gains, including higher household income (β = 198.649; 95% CI: 53.931, 343.366) and a higher probability of employment (β = 0.086; 95% CI: 0.004, 0.169). Some health-related behaviors and profiles also shift. College completion is associated with a lower likelihood of current smoking (β = −0.020, 95% CI: −0.035, −0.005) and fewer lifetime pregnancies (β = −0.209, 95% CI: −0.441, −0.023). We do not detect clear differences in vigorous physical activity, self-rated health, or the number of chronic conditions, and we cannot assess alcohol use or body mass index in this dataset. These patterns broadly mirror KNHANES, though estimates in KLoWF are generally less precise, likely reflecting its smaller sample from the narrower cohort window.

Additionally, college completion is associated with a 17.4 pp reduction in the weekly depressive-symptoms composite (β = −0.174, 95% CI: −0.576, −0.005) before adjusting for mediators. After adjusting for all mediators, the estimate attenuates only slightly to 16.4 pp (β = −0.164, 95% CI: −0.576, −0.005), indicating that these observed pathways account for little of the overall effect (**Table 5**).

**Robustness checks.** We re-estimated the models using placebo cohort cutoffs (1972, 1973, 1975, 1976) (**Table 11**, **Table 12**). Across these placebo samples, college completion shows no detectable effect on weekly depressive symptoms or on the daily affect/stress index (e.g., weekly depressive-symptoms composite: β = 1.083, 95% CI: −5.224, 7.390 in 1972; β = −0.002, 95% CI: −0.386, 0.382 in 1973; β = 0.923, 95% CI: −43.030, 44.875 in 1975; β = 0.581, 95% CI: −3.487, 4.648in 1976). The exceptionally wide intervals for some placebo years (especially 1975) are consistent with weak first-stage variation under these artificial thresholds. Overall, these placebo exercises reveal no systematic effects at the wrong cutoffs, supporting the timing and exclusion restrictions behind the main IV estimates.

## 5. Discussion

While prior research has documented strong associations between education and physical health, far less is known about the causal effects of higher education specifically. Moreover, limited research has evaluated impacts on mental health –particularly in non-Western contexts. Given the rising global burden of depression, especially among women in East Asia (50,69), understanding how higher education shapes mental well-being is both timely and policy-relevant.

This study examines the causal effect of college education on women's mental health in South Korea by leveraging the 1993 higher education reform as an exogenous source of variation in college attainment. Focusing on middle-aged women, we test whether expanded access to college reduces both clinically diagnosed depression and subjectively reported depressive symptoms, and assess mediation through economic conditions, health behaviors, and physical health. To bolster credibility and support a causal interpretation, we triangulate results across two independent datasets (KNHANES and KLoWF).

We find that college completion lowers the probability of a depression diagnosis by 2.4 pp, attenuating to 1.6 pp after adjusting for mediators. It is important to note that these estimates may understate the true effect, as we rely on self-reported physician diagnoses, which may miss subclinical or undiagnosed symptoms – especially among more economically disadvantaged populations with more limited access to health care. The protective effect appears to be partly mediated by improvements in economic conditions (household income and employment) and physical health (e.g., BMI, pregnancy history, and self-rated health), with only marginal attenuation observed through health behaviors such as smoking, drinking, and preventive care utilization. However, a substantial share of the effect remains unexplained, suggesting that other unmeasured pathways may also be at play. Placebo tests using unaffected cohorts yield null effects, and incorporating cohort-trend interactions does not attenuate the estimates, lending support to the exclusion restriction assumption in our IV strategy.

In parallel analyses of the KLoWF panel, instrumenting college completion with narrow birth-year windows around the 1974 cutoff, we also observe sizable improvements in self-reported mental well-being. Using the ±1-year window, college completion lowers the weekly depressive-symptoms composite by roughly 16 pp, and also reduces item-level measures (e.g., felt very depressed, difficulty concentrating). Results are directionally similar for the daily affect/stress index. Widening the bandwidth to ±2 and ±3 years yields larger reductions (≈19–31 pp) in weekly depressive symptoms; however, while these broader windows are more precise, they may introduce greater cohort heterogeneity. Additionally, robustness checks using placebo thresholds at adjacent birth years yield null effects, and first-stage F-statistics exceed conventional thresholds (>10), reinforcing identification. As in KNHANES, short-run changes in health behaviors or care use do not appear to explain the bulk of these improvements, pointing instead to broader channels through which higher education may buffer women's mental health.

Overall, the KNHANES and KLoWF results converge across distinct constructs: the reform-induced increase in women's college completion is associated with both a lower probability of diagnosed depression (KNHANES) and fewer depressive symptoms and adverse daily affect (KLoWF). Because diagnoses depend on care access and stigma, whereas symptoms capture experiences regardless of treatment seeking, this convergence strengthens a causal interpretation.

Our findings align with a broad literature – albeit primarily observational – documenting an inverse relationship between education and depression. For example, Li et al. (2021) reported significantly lower odds of depressive symptoms among college-educated adults in the U.S., though effects were weaker for minority and low-income groups (70). Similar patterns have been observed in China (71) and Belgium (72). A prior meta-analysis by Lorant et al. (2003) also confirmed a dose-response association, while also highlighting that the strength of socioeconomic inequalities in depression varies by measurement approach and contextual factors such as region and time (73). In South Korea, Lee (2011) documented educational gradients in depression and identified mechanisms including cognitive ability, economic resources, and social networks (74). Notably, these effects varied by age cohort: among younger adults (ages 40–55), economic conditions and social networks were stronger mediators, while cognitive ability played a larger role for older adults. Given that over 90% of our sample falls within the 40–55 age range, our findings regarding economic and behavioral pathways are consistent with Lee's results.

Our study also contributes to the small but growing set of quasi-experimental studies on education and mental health. McFarland and Wagner (2015), using a twin-difference design, found that higher education causally reduces depressive symptoms in the U.S (46). Similarly, Kondirolli and Sunder (2021) showed that a set of post-independence education reforms in Zimbabwe – including free and compulsory primary schooling and automatic progression to secondary education – led to lower rates of depression and anxiety, especially among women (75). Our findings extend this literature by focusing on college education–a relatively underexplored exposure in a quasi-experimental framework–and offering evidence from a non-Western, rapidly developing setting. While our effect sizes are somewhat smaller, potentially due to our focus on diagnosed depression rather than symptom severity, the results nonetheless support the mental health benefits of higher education.

Our findings are particularly salient in a sociocultural context like South Korea, where traditional gender norms and expectations have historically limited women's socioeconomic opportunities and reinforced a relational identity rooted in family roles (76). In such a setting, higher education may offer women more than just economic returns–it may also provide autonomy, psychological resilience, and access to supportive social networks. These findings underscore the importance of policies that reduce gender disparities in educational opportunities and support women's labor force participation and broader social integration.

However, our findings have several limitations. First, our study focuses on middle-aged women in South Korea, limiting the generalizability of findings to other populations, age groups, or male respondents. Second, although we use a quasi-experimental design, the potential for residual confounding remains – particularly from concurrent events surrounding the education reform that may independently affect both education and mental health. Third, we rely on a binary indicator of depression diagnosis, which may underestimate the true prevalence and severity of mental health conditions, especially among those who remain undiagnosed. Fourth, due to limited power and available measures, we could not explore subgroup differences (e.g., urban vs. rural) or heterogeneous effects. Future research should examine whether these effects vary across geography, family structure, caregiving roles, or intergenerational dynamics, that further shape women's mental health trajectories. Finally, while prior research suggests that cognitive ability may be an important mediator – particularly for older adults aged 55 and above (74) – we were unable to explore this pathway due to the lack of cognitive measures in our dataset. More nuanced mental health measures, including symptom severity and treatment access, would also enhance understanding of how education influences mental well-being across the life course.

## 6. Conclusion

Our findings offer new causal evidence that expanding access to college education can improve women's mental health in midlife, with reductions in depression diagnosis partially mediated through improved economic and physical well-being. These results underscore the mental health dividends of education policy and highlight the importance of targeting structural determinants – such as educational access–as part of a broader public health strategy to reduce gendered disparities in mental health.

# Tables and Figures

**Table 1. Association between 1993 Higher Education Reform and College Completion: 2SLS First-Stage Estimates**

|  | KNHANES (3-year birth cohort) | KLoWF (1-year birth cohort) |
|---|---|---|
|  | Beta [95% CI] | Beta [95% CI] |
| Exposure to the 1993 reform | 0.068*** | 0.116*** |
|  | (0.029 0.106) | (0.080 0.152) |
| N | 2912 | 2780 |
| F | 15 | 40 |

**Notes**. 1) The analysis draws on two datasets. For the Korea National Health and Nutrition Examination Survey (KNHANES), the sample includes women born in the three years after the 1993 higher-education reform (treatment group) and in the three years prior (control group). For the Korean Longitudinal Survey of Women and Families (KLoWF), we define treatment and control using birth-cohort windows around the 1974 cutoff (e.g., ±1-year: 1973 vs. 1974), where cohorts born in or after 1974 are considered exposed to the reform and earlier adjacent cohorts serve as controls. 2) All models adjust for respondent and family background, including age; marital status (if partnered); work status and occupation category; household characteristics (household size, number of siblings, and birth order [KLoWF only]); mother's education; birthplace (KLoWF only); household economic conditions in adolescence (KLoWF only); and parental relationship quality in adolescence (KLoWF only). (3) Coefficients are reported with 95% confidence intervals in brackets. Statistical significance: *** $p < 0.001$, ** $p < 0.01$, * $p < 0.05$.

**Table 2. Descriptive Statistics of Control and Treatment Groups (3-Year Cohorts) from Korea National Health and Nutrition Examination Survey (KNHANES; 2007–2021)**

|  | Control | Treatment | Test |
|---|---|---|---|
| N | 1,780 (61.1%) | 1,132 (38.9%) |  |
| Age, mean (sd) | 45.0 (2.9) | 43.0 (2.0) | <0.001*** |
| Father's education (unit: years), mean (sd) | 9.3 (4.1) | 9.6 (4.0) | 0.127 |
| Family history of chronic conditions (%) | 66.7% | 68.4% | 0.229 |
| Graduated college (%) | 47.5% | 56.0% | <0.001*** |
| Married (%) | 87.5% | 86.4% | 0.389 |
| Number of family members, mean (sd) | 3.5 (1.1) | 3.6 (1.1) | 0.003 |
| Monthly HH income (unit: 10,000 KRW) | 514.3 (297.1) | 546.1 (310.5) | <0.001*** |
| Current drinker | 75.8% | 78.0% | 0.155 |
| Current smoker | 6.0% | 5.4% | 0.441 |
| Received regular health check-up past two years | 72.9% | 75.5% | 0.116 |
| BMI | 23.2 (3.6) | 23.0 (3.7) | 0.112 |
| Number of pregnancies | 2.8 (1.5) | 2.7 (1.5) | <0.001*** |
| Diagnosis of depression (%) | 5.9% | 5.5% | 0.648 |

**Notes**. 1) The sample includes 1,132 individuals born in the three years following the 1993 Higher Education Reform (treatment group; born 1974-1966) and 1,780 individuals born in the three years prior (control group; born 1971-1973) from Korea National Health and Nutrition Examination Survey (KNHANES). 2) Family history of chronic conditions is an indicator equal to 1 if the respondent reports a family history of any of: hypertension, hyperlipidemia, ischemic heart disease (myocardial infarction or angina), stroke, diabetes, thyroid disease, or hepatitis B. 3) Statistical significance is denoted as ***p < 0.001, **p < 0.05, and *p < 0.1*.

**Table 3. Impact of College Education on Physician-Diagnosed Depression Incidence (KNHANES) and Self-reported Weekly Depressive Symptoms (KLoWF): 2SLS Second-Stage Estimates**

| Model | Description | Beta | 95% CI | N |
|---|---|---|---|---|
| **KNHANES (3-year birth cohort)** | | | | |
| M1 | Baseline model (no mediators) | -0.024*** | (-0.035 -0.012) | 2912 |
| M2 | M1 + Economic Conditions | -0.019*** | (-0.030 -0.008) | 2912 |
| M3 | M1 + Health Behaviors | -0.022*** | (-0.033 -0.011) | 2912 |
| M4 | M1 + Physical Health | -0.020*** | (-0.030 -0.009) | 2906 |
| M5 | M1 + All Mediators | -0.016*** | (-0.020 -0.003) | 2906 |
| **KLoWF (1-year birth cohort)** | | | | |
| M1 | Baseline model (no mediators) | -0.174** | (-0.576 -0.005) | 2780 |
| M2 | M1 + Economic Conditions | -0.162** | (-0.276 -0.049) | 2780 |
| M3 | M1 + Health Behaviors | -0.165** | (-0.279 -0.052) | 2780 |
| M4 | M1 + Physical Health | -0.169** | (-0.284 -0.055) | 2780 |
| M5 | M1 + All Mediators | -0.164** | (-0.576 -0.005) | 2780 |

**Notes**. 1) The analysis draws on two datasets. For the Korea National Health and Nutrition Examination Survey (KNHANES), the sample includes women born in the three years after the 1993 higher-education reform (treatment group) and in the three years prior (control group). For the Korean Longitudinal Survey of Women and Families (KLoWF), we define treatment and control using birth-cohort windows around the 1974 cutoff (e.g., ±1-year: 1973 vs. 1974), where cohorts born in or after 1974 are considered exposed to the reform and earlier adjacent cohorts serve as controls. 2) All models adjust for respondent and family background, including age; marital status (if partnered); work status and occupation category; household characteristics (household size, number of siblings, and birth order [KLoWF only]); mother's education; birthplace (KLoWF only); household economic conditions in adolescence (KLoWF only); and parental relationship quality in adolescence (KLoWF only). (3) Mediators are grouped as follows: Economic conditions– household income/assets, full-time employment status, and occupational category (both KNHANES and KLoWF); Health behaviors – smoking status, alcohol use, recent health check-up, and vigorous exercise (KNHANES includes smoking, alcohol, and check-up; KLoWF includes smoking and recent check-up); Physical health – body mass index, self-rated health, number of pregnancies, and number of chronic conditions (KNHANES includes BMI, self-rated health, and pregnancies; KLoWF includes self-rated health, pregnancies, and chronic conditions). 4) Coefficients are shown with 95% confidence intervals in brackets. Statistical significance: *** p < 0.001, ** p < 0.01, * p < 0.05.

Table 4. Impact of College Education on the Potential Mediators: 2SLS Second-Stage Estimates

| | | KNHANES (3-year birth cohort) | | | KLoWF (1-year birth cohort) | | |
|---|---|---|---|---|---|---|---|
| | Secondary outcomes | Beta | 95% CI | N | Beta | 95% CI | N |
| **Economic conditions** | Household income | 43.015*** | (27.331 58.700) | 2912 | 198.649** | (53.931 343.366) | 2780 |
| | Work status | 0.055 | (-0.048 0.158) | 2912 | 0.086* | (0.004 0.169) | 2780 |
| **Health behaviors** | Current drinker | 0.005 | (-0.012 0.022) | 2912 | n/a | n/a | n/a |
| | Current smoker | -0.014*** | (-0.023 -0.006) | 2912 | -0.020** | (-0.035 0.005) | 2780 |
| | Recent health check-up | 0.012 | (-0.008 0.033) | 2906 | 0.049 | (-0.225 0.323) | 2780 |
| | Vigorous exercise 3+ times in last 7 days | n/a | n/a | n/a | 0.091 | (-0.163 0.346) | 2780 |
| **Physical health conditions** | BMI | -0.355*** | (-0.523 -0.186) | 2906 | n/a | n/a | n/a |
| | Self-reported health | 0.012 | (-0.003 0.028) | 2912 | 0.131 | (-0.175 0.437) | 2780 |
| | Number of pregnancies | -0.160*** | (-0.222 -0.098) | 2912 | -0.209* | (-0.441 0.023) | 2780 |
| | Number of chronic conditions | n/a | n/a | n/a | 0.020 | (-0.077 0.117) | 2780 |

**Notes**. 1) The analysis draws on two datasets. For the Korea National Health and Nutrition Examination Survey (KNHANES), the sample includes women born in the three years after the 1993 higher-education reform (treatment group) and in the three years prior (control group). For the Korean Longitudinal Survey of Women and Families (KLoWF), we define treatment and control using birth-cohort windows around the 1974 cutoff (e.g., ±1-year: 1973 vs. 1974), where cohorts born in or after 1974 are considered exposed to the reform and earlier adjacent cohorts serve as controls. 2) All models adjust for respondent and family background, including age; marital status (if partnered); work status and occupation category; household characteristics (household size, number of siblings, and birth order [KLoWF only]); mother's education; birthplace (KLoWF only); household economic conditions in adolescence (KLoWF only); and parental relationship quality in adolescence (KLoWF only). (3) Mediators are grouped as follows: Economic conditions– household income/assets, full-time employment status, and occupational category (both KNHANES and KLoWF); Health behaviors – smoking status, alcohol use, recent health check-up, and vigorous exercise (KNHANES includes smoking, alcohol, and check-up; KLoWF includes smoking and recent check-up); Physical health – body mass index, self-rated health, number of pregnancies, and number of chronic conditions (KNHANES includes BMI, self-rated health, and pregnancies; KLoWF includes self-rated health, pregnancies, and chronic conditions). 4) Coefficients are shown with 95% confidence intervals in brackets. Statistical significance: *** p < 0.001, ** p < 0.01, * p < 0.05.

**Table 5. Placebo Test: Depression Outcomes Among Unexposed Cohorts (1968–1970 vs. 1971–1973)**

| | KNHANES (3-year birth cohort) | | | |
|---|---|---|---|---|
| Model | Description | Beta | 95% CI | N |
| M1 | Baseline model (no mediators) | -0.005 | (-0.014 0.004) | 4173 |
| M2 | M1 + Economic Conditions | -0.002 | (-0.011 0.008) | 4173 |
| M3 | M1 + Health Behaviors | -0.005 | (-0.013 0.004) | 4173 |
| M4 | M1 + Physical Health | -0.003 | (-0.012 0.005) | 4168 |
| M5 | M1 + All Mediators | -0.001 | (-0.007 0.009) | 4168 |

**Notes**. (1) The placebo sample includes women born 1964–1968 and 1969–1973, both unexposed to the 1993 higher-education reform, from the Korea National Health and Nutrition Examination Survey (KNHANES). The cohort indicator equals 1 for 1969–1973 and 0 for 1964–1968. (2) All models adjust for baseline covariates: age, marital status, occupation, household size, region of residence, and mother's education. (3) Model 1 (M1) includes only these baseline covariates; Model 2 (M2) adds economic conditions: household income, full-time employment status, and occupational category; Model 3 (M3) adds health behaviors: smoking status, alcohol consumption, and recent health check-up; Model 4 (M4) adds physical health indicators: body mass index (BMI), self-rated health, and number of pregnancies; Model 5 (M5) includes all mediators. (4) Coefficients are reported with 95% confidence intervals in brackets. Statistical significance is denoted as ***$p < 0.001$, **$p < 0.01$, and *$p < 0.05$.

**Table 6. Robustness Check: Impact of College Education on Depression with vs. without Cohort-Trend Interactions**

| Model | Description | Beta | 95% CI | N |
|---|---|---|---|---|
| M1 | Fully adjusted model without cohort-trend interactions | -0.016*** | (-0.020 -0.003) | 2906 |
| M2 | Fully adjusted model with cohort-trend interactions | -0.016*** | (-0.027 -0.005) | 2906 |

**Notes**. (1) The sample includes individuals born in the three years before and after the exposure window to the 1993 Higher Education Reform, all of whom were unaffected by the policy. Data are drawn from the Korea National Health and Nutrition Examination Survey (KNHANES). (2) Coefficients are derived from the second stage of 2SLS estimation and adjust for baseline covariates (age, marital status, occupation, household size, region of residence, and mother's education), economic conditions (household income, full-time employment status, occupational category), health behaviors (smoking, drinking, health check-up), and physical health indicators (BMI, self-rated health, number of pregnancies). (3) Coefficients are reported with 95% confidence intervals in brackets. Statistical significance is denoted as ***$p < 0.001$, **$p < 0.01$, and *$p < 0.05$.

**Table 7. Robustness Check: Alternative Birth Cohort Specifications**

| Model | Description | Beta | 95% CI | N |
|---|---|---|---|---|
| M1 | Fully adjusted model using 3-year birth cohort bandwidth | -0.016*** | (-0.027 -0.005) | 2906 |
| M2 | Fully adjusted model using 5-year birth cohort bandwidth (main model) | -0.009** | (-0.012 0.000) | 4985 |
| M3 | Fully adjusted model using 10-year birth cohort bandwidth | -0.003 | (-0.009 0.003) | 9718 |
| M4 | Fully adjusted model using 15-year birth cohort bandwidth | -0.004 | (-0.008 0.001) | 14946 |

**Notes**. (1) Models use alternative bandwidths for the birth cohort instrument to test robustness of the main results. (2) Data are drawn from the Korea National Health and Nutrition Examination Survey (KNHANES). (3) All models use 2SLS estimation and adjust for baseline covariates (age, marital status, occupation, household size, region, and mother's education), economic conditions (household income, full-time employment, occupation), health behaviors (smoking, drinking, health check-up), and physical health (BMI, self-rated health, number of pregnancies). (4) Coefficients are presented with 95% confidence intervals in parentheses. Statistical significance ***$p < 0.001$, **$p < 0.01$, and *$p < 0.05$.

**Table 8. Descriptive Statistics of Control and Treatment Groups: Korean Longitudinal Survey of Women and Families (KLoWF; 2007-2022) with 1-Year Cohort Window**

| | Exposure to 1993 higher-education reform | | | |
|---|---|---|---|---|
| | Control (1973) | Treated (1974) | Total | Test |
| N | 1386 (50%) | 1394 (50%) | 2780 (100%) | |
| Age | 40.73 (5.05) | 39.80 (5.05) | 40.27 (5.07) | <0.001 |
| Number of siblings | 3.86 (1.43) | 3.95 (1.53) | 3.90 (1.48) | 0.040 |
| Birthplace urbanicity (1=metro, 2=mid/small city, 3=rural, 4=overseas) | | | | |
|  1 Metro | 37% | 33% | 35% | <0.001 |
|  2 Mid/small city | 21% | 20% | 21% | |
|  3 Rural | 42% | 47% | 44% | |
| Father's education at age 15 | | | | |
|  Below high school | 78% | 80% | 79% | 0.106 |
|  High school or more | 22% | 20% | 21% | |
| Family financial situation at age 15 | 3.22 (0.79) | 3.25 (0.82) | 3.19 (0.81) | 0.211 |
| Parental relationship quality at age 15 | 2.83 (1.14) | 2.81 (1.25) | 2.74 (1.20) | 0.573 |
| Completed college or higher (2–3yr, 4yr, or graduate) | 0.45 (0.50) | 0.54 (0.50) | 0.50 (0.50) | <0.001 |
| Currently married | 0.89 (0.32) | 0.91 (0.28) | 0.90 (0.30) | 0.025 |
| Ever given birth | 1.83 (0.37) | 1.82 (0.38) | 1.83 (0.38) | 0.396 |
| Worked last week or not | 0.49 (0.50) | 0.52 (0.50) | 0.50 (0.50) | 0.030 |
| Ever smoked | 2.97 (0.21) | 2.98 (0.18) | 2.98 (0.19) | 0.507 |
| Had a regular health checkup | 1.24 (0.43) | 1.27 (0.44) | 1.25 (0.43) | 0.048 |
| Used any medical service in past year | 1.48 (0.50) | 1.55 (0.50) | 1.52 (0.50) | <0.001 |

**Notes**. (1) The sample includes 382 individuals born in 1974 – the first cohort exposed to the 1993 Higher Education Reform (treatment) – and 391 individuals born in 1973 (control) from the Korean Longitudinal Survey of Women and Families (KLoWF). (2) Descriptive statistics for wider cohort windows (±2 years and ±3 years) are reported in eTable 1 and eTable 2. (3) Statistical significance: *** $p < 0.001$, ** $p < 0.01$, * $p < 0.05$.

**Table 9. Impact of College Education on Weekly Depressive Symptoms: 2SLS Second-Stage Estimates (KLoWF) with 1-Year Cohort Window**

|  | Beta | 95% CI | N |
|---|---|---|---|
| Felt bothered | -0.221* | (-0.866 0.034) | 2780 |
| Difficulty concentrating | -0.206** | (-0.805 0.001) | 2780 |
| Felt very depressed | -0.200* | (-0.805 0.008) | 2780 |
| Found things very hard | -0.154* | (-0.672 0.198) | 2780 |
| Felt I did okay (reverse coded) | -0.152 | (-1.028 0.006) | 2780 |
| Felt afraid | -0.141* | (-0.449 0.156) | 2780 |
| Trouble sleeping | -0.110 | (-0.462 0.294) | 2780 |
| Had no major complaints (reverse coded) | -0.115 | (-0.729 0.375) | 2780 |
| Felt lonely | -0.187* | (-0.760 0.015) | 2780 |
| Had no energy | -0.149* | (-0.466 0.149) | 2780 |
| **Depressive symptoms composite** | **-0.164**** | **(-0.576 -0.005)** | **2780** |

Notes. (1) The ±1-year window compares women born in 1973 (control) to those born in 1974 (treated via exposure to the 1993 Higher Education Reform), using the Korean Longitudinal Survey of Women and Families (KLoWF). (2) Entries are second-stage 2SLS coefficients for college completion on each binary symptom indicator (1 = endorses the adverse state; reverse-coded where noted) and on the composite index (mean of the ten binaries; range 0–1). (3) Covariates include: respondent age; birth order; number of siblings; birthplace (indicator set for 1 = metro, 2 = mid/small city, 3 = rural, 4 = overseas); mother's education at age 15; household economic condition at age 15; and parents' relationship quality at age 15. (4) Coefficients are reported with 95% confidence intervals in brackets. Statistical significance: *** $p < 0.001$, ** $p < 0.01$, * $p < 0.05$.

**Table 10. Impact of College Education on Daily Affects/ Stress: 2SLS Second-Stage Estimates with 1-Year Cohort Window (KLoWF)**

| 1 Year | Beta | 95% CI | N |
|---|---|---|---|
| Felt stressed | 0.015 | (-0.966 0.407) | 2780 |
| Drank alone due to stress | -0.107 | (-0.684 0.129) | 2780 |
| Preferred solitude | -0.129 | (-1.085 0.082) | 2780 |
| Got angry | -0.120 | (-0.540 0.204) | 2780 |
| No one to talk to | -0.223* | (-0.973 0.081) | 2780 |
| Managed stress well (reverse coded) | -0.406* | (-0.972 0.396) | 2780 |
| Financial stress | -0.223 | (-1.302 0.110) | 2780 |
| Interpersonal stress | -0.112 | (-0.885 0.370) | 2780 |
| **Daily affect/stress index** | **-0.163*** | **(-0.691 -0.012)** | **2780** |

Notes. (1) The ±1-year window compares women born in 1973 (control) to those born in 1974 (treated via exposure to the 1993 Higher Education Reform), using the Korean Longitudinal Survey of Women and Families (KLoWF). (2) Entries are second-stage 2SLS coefficients for college completion on each daily affect/emotion item (coded 1 if the adverse state is endorsed; "managed stress well" is reverse-coded so higher values indicate worse affect) and on the composite index (mean of the eight binary items; range 0–1). (3) Covariates include: respondent age; birth order; number of siblings; birthplace (1 = metro, 2 = mid/small city, 3 = rural, 4 = overseas); mother's education at age 15; household economic condition at age 15; parents' relationship quality at age 15; health behaviors (current smoking, recent preventive check-up); economic conditions (household assets, full-time employment status, and occupational category); and physical health indicators (self-rated health, number of pregnancies, and number of chronic conditions). (4) Coefficients are reported with 95% confidence intervals in brackets. Statistical significance: *** $p < 0.001$, ** $p < 0.01$, * $p < 0.05$.

**Table 11. Impact of College Education on Weekly Depressive-Symptoms Composite: 2SLS Second-Stage Estimates with 1-Year Cohort Window (KLoWF) and Placebo Cutoffs (1972, 1973, 1975, and 1976)**

| Placebo Dates | Beta | 95% CI | N |
|---|---|---|---|
| 1972 | 0.556 | (-3.008 4.119) | 3056 |
| 1973 | -0.104 | (-0.636 0.428) | 2863 |
| 1975 | 9.707 | (-433.631 453.045) | 2811 |
| 1976 | 0.370 | (-2.723 3.464) | 2478 |

**Notes**. (1) Each panel uses a ±1-year cohort window centered on the listed placebo cutoff. For 1972 and 1973, both cohorts are pre-reform (neither is exposed). For 1975 and 1976, the post-cutoff cohort is post-1974 and thus exposed to the reform, but these are not the first cohorts affected, so any exposure contrast is expected to be smaller than at the true 1974 threshold. The instrument is an indicator for being in the post-cutoff cohort; estimates are from the Korean Longitudinal Survey of Women and Families (KLoWF). (2) Entries are second-stage 2SLS coefficients for college completion on the weekly depressive-symptoms composite (mean of the ten binary items; range 0–1). Item-level estimates are reported in eTable 6. (3) Covariates include: respondent age; birth order; number of siblings; birthplace (1 = metro, 2 = mid/small city, 3 = rural, 4 = overseas); mother's education at age 15; household economic condition at age 15; parents' relationship quality at age 15; health behaviors (current smoking, recent preventive check-up); economic conditions (household assets, full-time employment status, and occupational category); and physical health indicators (self-rated health, number of pregnancies, and number of chronic conditions). (4) Coefficients are reported with 95% confidence intervals in brackets. Statistical significance: *** $p < 0.001$, ** $p < 0.01$, * $p < 0.05$.

**Table 12. Impact of College Education on Daily Affects/Stress Composite: 2SLS Second-Stage Estimates with 1-Year Cohort Window (KLoWF) and Placebo Cutoffs (1972, 1973, 1975, and 1976)**

| Placebo Dates | Beta | 95% CI | N |
|---|---|---|---|
| 1972 | 1.083 | (-5.224 7.390) | 3056 |
| 1973 | -0.002 | (-0.386 0.382) | 2863 |
| 1975 | 0.923 | (-43.030 44.875) | 2811 |
| 1976 | 0.581 | (-3.487 4.648) | 2478 |

**Notes**. (1) Each panel uses a ±1-year cohort window centered on the listed placebo cutoff. For 1972 and 1973, both cohorts are pre-reform (neither is exposed). For 1975 and 1976, the post-cutoff cohort is post-1974 and thus exposed to the reform, but these are not the first cohorts affected, so any exposure contrast is expected to be smaller than at the true 1974 threshold. The instrument is an indicator for being in the post-cutoff cohort; estimates are from the Korean Longitudinal Survey of Women and Families (KLoWF). (2) Entries are second-stage 2SLS coefficients for college completion on the daily affects/emotions composite (mean of the eight binary items; range 0–1). Item-level estimates are reported in eTable 7. (3) Covariates include: respondent age; birth order; number of siblings; birthplace (1 = metro, 2 = mid/small city, 3 = rural, 4 = overseas); mother's education at age 15; household economic condition at age 15; parents' relationship quality at age 15; health behaviors (current smoking, recent preventive check-up); economic conditions (household assets, full-time employment status, and occupational category); and physical health indicators (self-rated health, number of pregnancies, and number of chronic conditions). (4) Coefficients are reported with 95% confidence intervals in brackets. Statistical significance: *** $p < 0.001$, ** $p < 0.01$, * $p < 0.05$.

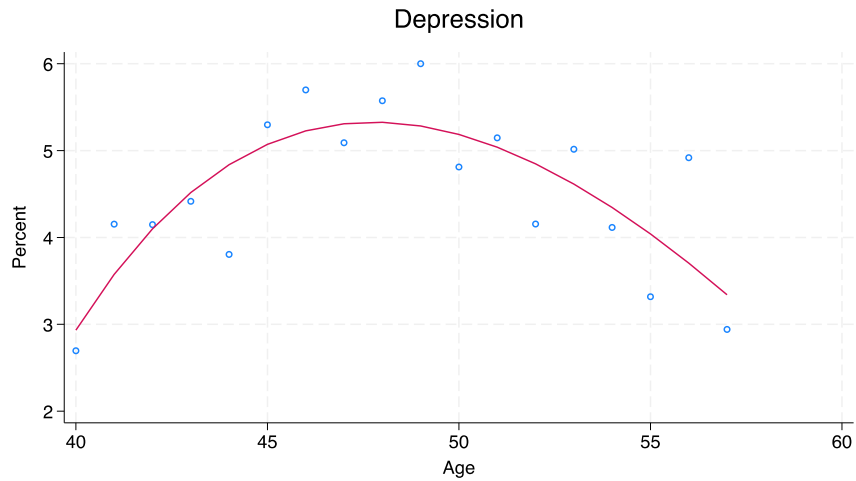

**Figure 1. Age Pattern of Depression Prevalence Among Middle-Aged Women in South Korea**
**Note**: This figure shows the prevalence of diagnosed depression among female respondents (aged 40-65) from the Korea National Health and Nutrition Examination Survey (KNHANES).

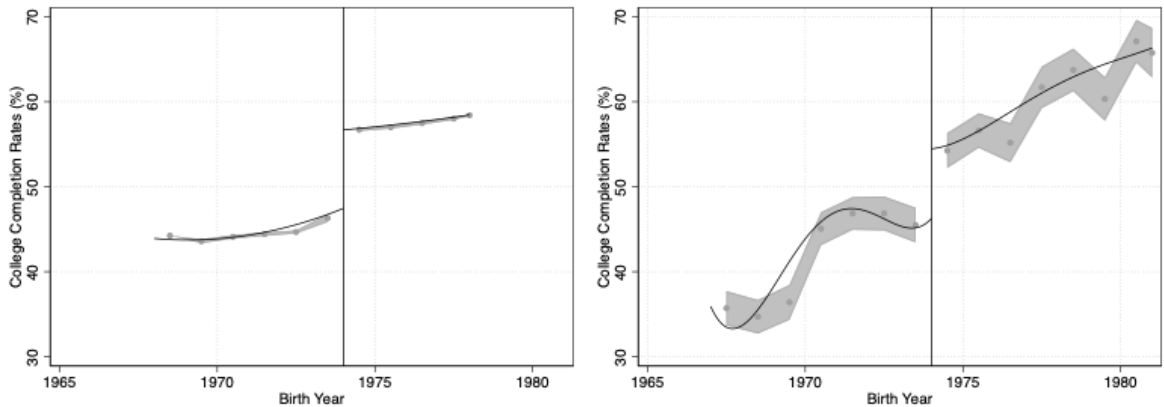

**Figure 2. Trends in Women's College Completion Rates by Birth Cohort in South Korea**
**Note**: The figures show college completion rates of the female respondents from the Korea National Health and Nutrition Examination Survey (KNHANES; left panel) and the Korean Longitudinal Survey of Women and Families (KLoWF; right panel). Both panels display cohorts on either side of the 1974 cutoff – the first birth cohort affected by the 1993 higher-education reform.

# Supplementary Information

**eTable 1. Descriptive Statistics of Control and Treatment Groups: Korean Longitudinal Survey of Women and Families (KLoWF; 2007-2022) with 2-Year Cohort Window**

|  | Exposure to 1993 higher-education reform | | | |
|---|---|---|---|---|
|  | Treated (1974-1975) | Control (1972-1973) | Total | Test |
| N | 2863 (50%) | 2811 (50%) | 5674 (100%) |  |
| Age | 39.27 (5.07) | 41.24 (5.08) | 40.27 (5.17) | <0.001 |
| Number of siblings | 3.82 (1.49) | 3.93 (1.50) | 3.87 (1.50) | <0.001 |
| Birthplace urbanicity (1=metro, 2=mid/small city, 3=rural, 4=overseas) |  |  |  |  |
|   1 Metro | 33% | 34% | 34% | 0.208 |
|   2 Mid/small city | 22% | 21% | 21% |  |
|   3 Rural | 44% | 45% | 45% |  |
| Mother's education at age 15 |  |  |  |  |
|   Below high school | 79% | 80% | 80% | 0.152 |
|   High school or more | 21% | 20% | 20% |  |
| Family financial situation at age 15 | 3.19 (0.79) | 3.14 (0.82) | 3.17 (0.80) | 0.005 |
| Parental relationship quality at age 15 | 2.82 (1.21) | 2.76 (1.25) | 2.79 (1.23) | 0.030 |
| Completed college or higher (2–3yr, 4yr, or graduate) | 0.55 (0.50) | 0.46 (0.50) | 0.51 (0.50) | <0.001 |
| Currently married | 0.89 (0.31) | 0.90 (0.29) | 0.90 (0.30) | 0.110 |
| Ever given birth | 1.82 (0.38) | 1.83 (0.37) | 1.83 (0.38) | 0.064 |
| Worked last week or not | 0.49 (0.50) | 0.50 (0.50) | 0.49 (0.50) | 0.223 |
| Ever smoked | 2.97 (0.21) | 2.98 (0.18) | 2.98 (0.19) | 0.191 |
| Had a regular health checkup | 1.28 (0.45) | 1.23 (0.42) | 1.26 (0.44) | <0.001 |
| Used any medical service in past year | 1.56 (0.50) | 1.50 (0.50) | 1.53 (0.50) | <0.001 |

**Notes**. (1) The sample includes 770 individuals born in 1974-1975 – the first two cohorts exposed to the 1993 Higher Education Reform (treatment) – and 784 individuals born in 1972–1973 (control) from the Korean Longitudinal Survey of Women and Families (KLoWF). (2) Statistical significance: *** $p < 0.001$, ** $p < 0.01$, * $p < 0.05$.

**eTable 2. Descriptive Statistics of Control and Treatment Groups: Korean Longitudinal Survey of Women and Families (KLoWF; 2007-2022) with 3-Year Cohort Window**

| | Exposure to 1993 higher-education reform | | | |
|---|---|---|---|---|
| | Treated (1974-1976) | Control (1971-1973) | Total | Test |
| N | 4442 (53%) | 3872 (47%) | 8314 (100%) | |
| Age | 38.82 (5.11) | 41.75 (5.12) | 40.39 (5.32) | <0.001 |
| Number of siblings | 3.74 (1.44) | 3.98 (1.51) | 3.87 (1.48) | <0.001 |
| Birthplace urbanicity (1=metro, 2=mid/small city, 3=rural, 4=overseas) | | | | |
| 1 Metro | 33% | 35% | 34% | <0.001 |
| 2 Mid/small city | 24% | 20% | 22% | |
| 3 Rural | 44% | 45% | 44% | |
| Mother's education at age 15 | | | | |
| Below high school | 78% | 81% | 79% | <0.001 |
| High school or more | 22% | 19% | 21% | |
| Family financial situation at age 15 | 3.18 (0.79) | 3.17 (0.82) | 3.17 (0.80) | 0.798 |
| Parental relationship quality at age 15 | 2.77 (1.22) | 2.75 (1.23) | 2.76 (1.23) | 0.287 |
| Completed college or higher (2–3yr, 4yr, or graduate) | 0.55 (0.50) | 0.46 (0.50) | 0.51 (0.50) | <0.001 |
| Currently married | 0.90 (0.30) | 0.90 (0.30) | 0.90 (0.30) | 0.523 |
| Ever given birth | 1.82 (0.39) | 1.83 (0.37) | 1.83 (0.38) | 0.007 |
| Worked last week or not | 0.48 (0.50) | 0.52 (0.50) | 0.50 (0.50) | <0.001 |
| Ever smoked | 2.97 (0.21) | 2.97 (0.20) | 2.97 (0.21) | 0.564 |
| Had a regular health checkup | 1.28 (0.45) | 1.23 (0.42) | 1.26 (0.44) | <0.001 |
| Used any medical service in past year | 1.54 (0.50) | 1.51 (0.50) | 1.52 (0.50) | <0.001 |

**Notes**. (1) The sample includes 1,231 individuals born in 1974-1976 – the first three cohorts exposed to the 1993 Higher Education Reform (treatment) – and 1,086 individuals born in 1971–1973 (control) from the Korean Longitudinal Survey of Women and Families (KLoWF). (2) Statistical significance: *** p < 0.001, ** p < 0.01, * p < 0.05.

**eTable 3. Impact of College Education on Weekly Depressive Symptoms: 2SLS Second-Stage Estimates (KLoWF) with 2- and 3-Year Cohort Windows**

| 2 Year | Beta | 95% CI | N |
|---|---|---|---|
| Felt bothered | -0.339*** | (-0.523 -0.156) | 5674 |
| Difficulty concentrating | -0.173** | (-0.303 -0.043) | 5674 |
| Felt very depressed | -0.209** | (-0.353 -0.065) | 5674 |
| Found things very hard | -0.135* | (-0.284 0.015) | 5674 |
| Felt I did okay (reverse coded) | -0.178* | (-0.367 0.012) | 5674 |
| Felt afraid | -0.132* | (-0.248 -0.015) | 5674 |
| Trouble sleeping | -0.081 | (-0.223 0.061) | 5674 |
| Had no major complaints (reverse coded) | -0.342** | (-0.587 -0.096) | 5674 |
| Felt lonely | -0.186** | (-0.319 -0.053) | 5674 |
| Had no energy | -0.164** | (-0.287 -0.041) | 5674 |
| **Depressive symptoms composite** | **-0.194*** | **(-0.297 -0.091)** | 5674 |
| 2 Year | Beta | 95% CI | N |
| Felt bothered | -0.532*** | (-0.751 -0.314) | 8550 |
| Difficulty concentrating | -0.245*** | (-0.387 -0.103) | 8550 |
| Felt very depressed | -0.315*** | (-0.476 -0.154) | 8550 |
| Found things very hard | -0.246** | (-0.410 -0.082) | 8550 |
| Felt I did okay (reverse coded) | -0.271** | (-0.474 -0.069) | 8550 |
| Felt afraid | -0.145* | (-0.265 -0.025) | 8550 |
| Trouble sleeping | -0.122 | (-0.269 0.025) | 8550 |
| Had no major complaints (reverse coded) | -0.714*** | (-1.018 -0.409) | 8550 |
| Felt lonely | -0.288*** | (-0.438 -0.138) | 8550 |
| Had no energy | -0.225*** | (-0.355 -0.094) | 8550 |
| **Depressive symptoms composite** | **-0.310*** | **(-0.434 -0.187)** | 8550 |

**Notes**. (1) The ±2-year window compares women born 1972–1973 (control) to 1974–1975 (treated via exposure to the 1993 Higher Education Reform); the ±3-year window compares 1971–1973 (control) to 1974–1976 (treated), using the Korean Longitudinal Survey of Women and Families (KLoWF). (2) Entries are second-stage 2SLS coefficients for college completion on each binary symptom indicator (1 = endorses the adverse state; reverse-coded where noted) and on the composite index (mean of the ten binaries; range 0–1). (3) Covariates include: respondent age; birth order; number of siblings; birthplace (indicator set for 1 = metro, 2 = mid/small city, 3 = rural, 4 = overseas); mother's education at age 15; household economic condition at age 15; and parents' relationship quality at age 15. (4) Coefficients are reported with 95% confidence intervals in brackets. Statistical significance: *** p < 0.001, ** p < 0.01, * p < 0.05.

**eTable 4. Impact of College Education on Daily Affects/Emotions: 2SLS Second-Stage Estimates (KLoWF) with 2- and 3-Year Cohort Windows**

| 2 Year Cohort Window | Beta | 95% CI | N |
|---|---|---|---|
| Felt stressed | -0.197 | (-0.490 0.097) | 5674 |
| Drank alone due to stress | -0.035 | (-0.206 0.135) | 5674 |
| Preferred solitude | -0.071 | (-0.268 0.126) | 5674 |
| Got angry | -0.119 | (-0.286 0.049) | 5674 |
| No one to talk to | 0.057 | (-0.145 0.260) | 5674 |
| Managed stress well (reverse coded) | -0.099 | (-0.388 0.191) | 5674 |
| Financial stress | -0.274* | (-0.560 0.011) | 5674 |
| Interpersonal stress | 0.027 | (-0.245 0.300) | 5674 |
| **Daily affect/stress index** | **-0.089** | **(-0.210 0.033)** | **5674** |
| **3 Year Cohort Window** | **Beta** | **95% CI** | **N** |
| Felt stressed | -0.388* | (-0.702 -0.073) | 8550 |
| Drank alone due to stress | -0.103 | (-0.278 0.073) | 8550 |
| Preferred solitude | 0.011 | (-0.191 0.213) | 8550 |
| Got angry | -0.147* | (-0.318 0.024) | 8550 |
| No one to talk to | 0.008 | (-0.197 0.213) | 8550 |
| Managed stress well (reverse coded) | -0.027 | (-0.320 0.266) | 8550 |
| Financial stress | -0.493** | (-0.801 -0.186) | 8550 |
| Interpersonal stress | -0.009 | (-0.287 0.269) | 8550 |
| **Daily affect/stress index** | **-0.143*** | **(-0.270 -0.017)** | **8550** |

**Notes**. (1) The ±2-year window compares women born 1972–1973 (control) to 1974–1975 (treated via exposure to the 1993 Higher Education Reform); the ±3-year window compares 1971–1973 (control) to 1974–1976 (treated), using the Korean Longitudinal Survey of Women and Families (KLoWF). (2) Entries are second-stage 2SLS coefficients for college completion on each binary symptom indicator (1 = endorses the adverse state; reverse-coded where noted) and on the composite index (mean of the eight binaries; range 0–1). (3) Covariates include: respondent age; birth order; number of siblings; birthplace (indicator set for 1 = metro, 2 = mid/small city, 3 = rural, 4 = overseas); mother's education at age 15; household economic condition at age 15; and parents' relationship quality at age 15. (4) Coefficients are reported with 95% confidence intervals in brackets. Statistical significance: *** $p < 0.001$, ** $p < 0.01$, * $p < 0.05$.

**eTable 5. Impact of College Education on Potential Mediators: 2SLS Second-Stage Estimates (KLoWF) with 2- and 3-Year Cohort Windows**

| 2 Year Cohort Window | Beta | 95% CI | N |
|---|---|---|---|
| Used any medical service in past year | -0.063 | (-0.348 0.222) | 5674 |
| Had a regular health checkup | -0.221* | (-0.471 0.029) | 5674 |
| Ever smoked | -0.039 | (-0.148 0.070) | 5674 |
| Vigorous exercise 3+ times in last 7 days | 0.036 | (-0.191 0.263) | 5674 |
| **3 Year Cohort Window** | **Beta** | **95% CI** | **N** |
| Used any medical service in past year | -0.544** | (-0.870 -0.218) | 8550 |
| Had a regular health checkup | -0.531*** | (-0.815 -0.248) | 8550 |
| Ever smoked | 0.001 | (-0.117 0.118) | 8550 |
| Vigorous exercise 3+ times in last 7 days | 0.236* | (-0.005 0.477) | 8550 |

**Notes**. (1) The ±2-year window compares women born 1972–1973 (control) to 1974–1975 (treated via exposure to the 1993 Higher Education Reform); the ±3-year window compares 1971–1973 (control) to 1974–1976 (treated), using the Korean Longitudinal Survey of Women and Families (KLoWF). (2) Entries are second-stage 2SLS coefficients for college completion on each mediator: any medical service use in the past year; regular health checkup in the past two years; ever smoked; and vigorous physical activity 3+ days in the last 7 days (all coded as binary indicators). (3) Covariates include: respondent age; birth order; number of siblings; birthplace (indicator set for 1 = metro, 2 = mid/small city, 3 = rural, 4 = overseas); mother's education at age 15; household economic condition at age 15; and parents' relationship quality at age 15. (4) Coefficients are reported with 95% confidence intervals in brackets. Statistical significance: *** p < 0.001, ** p < 0.01, * p < 0.05.

**eTable 6. Impact of College Education on Daily Affects/Emotions:
2SLS Second-Stage Estimates with 1-Year Cohort Window (KLoWF) and Placebo Cutoffs
(1972, 1973, 1975, and 1976)**

| 1972 | Beta | 95% CI | N |
|---|---|---|---|
| Felt bothered | 0.707 | (-3.647 5.061) | 3056 |
| Difficulty concentrating | 0.481 | (-2.539 3.501) | 3056 |
| Felt very depressed | 0.647 | (-3.373 4.667) | 3056 |
| Found things very hard | 1.212 | (-5.878 8.303) | 3056 |
| Felt I did okay (reverse coded) | 1.483 | (-7.461 10.428) | 3056 |
| Felt afraid | 0.177 | (-1.340 1.695) | 3056 |
| Trouble sleeping | 1.256 | (-6.017 8.529) | 3056 |
| Had no major complaints (reverse coded) | 4.165 | (-19.733 28.063) | 3056 |
| Felt lonely | 0.556 | (-2.938 4.050) | 3056 |
| Had no energy | 0.144 | (-1.298 1.586) | 3056 |
| **Depressive symptoms composite** | **1.083** | **(-5.224 7.390)** | **3056** |
| **1973** | **Beta** | **95% CI** | **N** |
| Felt bothered | 0.355 | (-0.552 1.261) | 2863 |
| Difficulty concentrating | -0.133 | (-0.646 0.380) | 2863 |
| Felt very depressed | 0.087 | (-0.429 0.604) | 2863 |
| Found things very hard | -0.158 | (-0.724 0.408) | 2863 |
| Felt I did okay (reverse coded) | 0.027 | (-0.691 0.745) | 2863 |
| Felt afraid | -0.130 | (-0.564 0.304) | 2863 |
| Trouble sleeping | -0.213 | (-0.824 0.398) | 2863 |
| Had no major complaints (reverse coded) | 0.200 | (-0.858 1.258) | 2863 |
| Felt lonely | 0.024 | (-0.455 0.503) | 2863 |
| Had no energy | -0.082 | (-0.518 0.353) | 2863 |
| **Depressive symptoms composite** | **-0.002** | **(-0.386 0.382)** | **2863** |
| **1975** | **Beta** | **95% CI** | **N** |
| Felt bothered | 3.117 | (-139.972 146.206) | 2811 |
| Difficulty concentrating | 4.790 | (-214.730 224.310) | 2811 |
| Felt very depressed | 5.802 | (-259.673 271.278) | 2811 |
| Found things very hard | 1.432 | (-65.758 68.621) | 2811 |
| Felt I did okay (reverse coded) | 0.672 | (-35.551 36.896) | 2811 |
| Felt afraid | 2.695 | (-121.458 126.848) | 2811 |
| Trouble sleeping | 1.884 | (-86.167 89.935) | 2811 |
| Had no major complaints (reverse coded) | -17.339 | (-806.736 772.058) | 2811 |
| Felt lonely | 5.899 | (-264.230 276.029) | 2811 |
| Had no energy | 0.276 | (-16.160 16.712) | 2811 |
| **Depressive symptoms composite** | **0.923** | **(-43.030 44.875)** | **2811** |
| **1976** | **Beta** | **95% CI** | **0** |
| Felt bothered | 2.213 | (-12.231 16.657) | 2478 |
| Difficulty concentrating | 1.021 | (-5.867 7.908) | 2478 |
| Felt very depressed | 0.915 | (-5.244 7.075) | 2478 |
| Found things very hard | 0.836 | (-4.887 6.559) | 2478 |
| Felt I did okay (reverse coded) | -0.724 | (-5.946 4.497) | 2478 |
| Felt afraid | 0.249 | (-1.858 2.356) | 2478 |
| Trouble sleeping | -0.413 | (-3.455 2.629) | 2478 |
| Had no major complaints (reverse coded) | 0.427 | (-4.000 4.854) | 2478 |
| Felt lonely | 0.628 | (-3.831 5.086) | 2478 |
| Had no energy | 0.657 | (-3.906 5.219) | 2478 |
| **Depressive symptoms composite** | **0.581** | **(-3.487 4.648)** | **2478** |

**Notes**. (1) Each panel uses a ±1 cohort window centered on the listed placebo cutoff. For 1972 and 1973, both cohorts are pre-reform (neither is exposed). For 1975 and 1976, the post-cutoff cohort is post-1974 and thus exposed to the reform, but these are not the first cohorts affected, so any exposure contrast is expected to be smaller than at the true 1974 threshold. The instrument is an indicator for being in the post-cutoff cohort; estimates are from the Korean Longitudinal Survey of Women and Families (KLoWF). (2) Entries are second-stage 2SLS coefficients for college completion on the weekly depressive-symptoms. (3) Covariates include: respondent age; birth order; number of siblings; birthplace (indicator set for 1 = metro, 2 = mid/small city, 3 = rural, 4 = overseas); mother's education at age 15; household economic condition at age 15; and parents' relationship quality at age 15. (4) Coefficients are reported with 95% confidence intervals in brackets. Statistical significance: *** p < 0.001, ** p < 0.01, * p < 0.05.

**eTable 7. Impact of College Education on Weekly Depressive-Symptoms: 2SLS Second-Stage Estimates with 1-Year Cohort Window (KLoWF) and Placebo Cutoffs (1972, 1973, 1975, and 1976)**

| 1972 | Beta | 95% CI | N |
|---|---|---|---|
| Felt stressed | -0.022 | (-3.362 3.318) | 3056 |
| Drank alone due to stress | 1.632 | (-7.822 11.086) | 3056 |
| Preferred solitude | 0.879 | (-4.608 6.365) | 3056 |
| Got angry | -0.118 | (-2.024 1.789) | 3056 |
| No one to talk to | 0.827 | (-4.550 6.204) | 3056 |
| Managed stress well (reverse coded) | -0.921 | (-6.934 5.092) | 3056 |
| Financial stress | 1.222 | (-6.897 9.342) | 3056 |
| Interpersonal stress | 0.945 | (-5.315 7.204) | 3056 |
| **Daily affect/stress index** | **0.556** | **(-3.008 4.119)** | **3056** |
| **1973** | **Beta** | **95% CI** | **N** |
| Felt stressed | 0.805 | (-1.065 2.676) | 2863 |
| Drank alone due to stress | -0.252 | (-1.056 0.552) | 2863 |
| Preferred solitude | -0.060 | (-0.780 0.660) | 2863 |
| Got angry | 0.228 | (-0.533 0.990) | 2863 |
| No one to talk to | -0.466 | (-1.585 0.653) | 2863 |
| Managed stress well (reverse coded) | -1.074 | (-3.394 1.246) | 2863 |
| Financial stress | 0.147 | (-1.156 1.450) | 2863 |
| Interpersonal stress | -0.161 | (-1.245 0.924) | 2863 |
| **Daily affect/stress index** | **-0.104** | **(-0.636 0.428)** | **2863** |
| **1975** | **Beta** | **95% CI** | **N** |
| Felt stressed | -1.667 | (-85.562 82.227) | 2811 |
| Drank alone due to stress | 3.843 | (-173.162 180.848) | 2811 |
| Preferred solitude | 10.453 | (-466.759 487.664) | 2811 |
| Got angry | 10.151 | (-453.170 473.472) | 2811 |
| No one to talk to | 20.307 | (-906.309 946.923) | 2811 |
| Managed stress well (reverse coded) | 14.496 | (-648.756 677.748) | 2811 |
| Financial stress | 5.985 | (-271.640 283.610) | 2811 |
| Interpersonal stress | 14.089 | (-628.862 657.040) | 2811 |
| **Daily affect/stress index** | **9.707** | **(-433.631 453.045)** | **2811** |
| **1976** | **Beta** | **95% CI** | **N** |
| Felt stressed | 1.591 | (-9.329 12.510) | 2478 |
| Drank alone due to stress | -0.064 | (-2.437 2.310) | 2478 |
| Preferred solitude | -2.417 | (-18.206 13.372) | 2478 |
| Got angry | 0.634 | (-4.013 5.282) | 2478 |
| No one to talk to | 1.110 | (-6.828 9.048) | 2478 |
| Managed stress well (reverse coded) | 0.857 | (-6.044 7.758) | 2478 |
| Financial stress | 1.810 | (-11.176 14.797) | 2478 |
| Interpersonal stress | -0.560 | (-5.835 4.715) | 2478 |
| **Daily affect/stress index** | **0.370** | **(-2.723 3.464)** | **2478** |

**Notes**. (1) Each panel uses a ±1 cohort window centered on the listed placebo cutoff. For 1972 and 1973, both cohorts are pre-reform (neither is exposed). For 1975 and 1976, the post-cutoff cohort is post-1974 and thus exposed to the reform, but these are not the first cohorts affected, so any exposure contrast is expected to be smaller than at the true 1974 threshold. The instrument is an indicator for being in the post-cutoff cohort; estimates are from the Korean Longitudinal Survey of Women and Families (KLoWF). (2) Entries are second-stage 2SLS coefficients for college completion on the daily affects/emotions. (3) Covariates include: respondent age; birth order; number of siblings; birthplace (indicator set for 1 = metro, 2 = mid/small city, 3 = rural, 4 = overseas); mother's education at age 15; household economic condition at age 15; and parents' relationship quality at age 15. (4) Coefficients are reported with 95% confidence intervals in brackets. Statistical significance: *** $p < 0.001$, ** $p < 0.01$, * $p < 0.05$.

**Appendix 1 Variable Definitions**

**1. Korea National Health and Nutrition Examination Survey (KNHANES)**

*Economic conditions* Household income is measured as a continuous variable (in units of 10,000 KRW), adjusted for inflation. Full-time employment status is measured as a binary indicator (employed full-time vs. not). Occupational status is categorized into seven groups: (1) managers, professionals, and related workers, (2) clerical workers, (3) service and sales workers, (4) skilled agricultural, forestry, and fishery workers, (5) craft and machine operators and assemblers, (6) elementary occupations (e.g., manual labor), and (7) those not in the labor force (e.g., homemakers, students).

*Health behaviors* Smoking status is measured as a binary variable, distinguishing current smokers (daily or occasional) from former or non-smokers. Drinking status is similarly categorized, identifying individuals who consume alcohol at least occasionally versus those who abstain. Engagement in preventive care is captured by whether respondents received a health check-up in the past two years.

*Physical health* Self-reported health is based on a five-point Likert scale and dichotomized into "good" (very good, good, or average) versus "poor" (poor or very poor). BMI is based on physical measurements of height and weight and is included as a continuous variable to capture variation in weight-related health outcomes. Pregnancy experience is measured as a count of lifetime pregnancies, given its relevance to midlife physical and mental health among women (65). Although physical activity was a mediator of interest, inconsistencies in data collection across survey waves precluded its inclusion.

**2. Korean Longitudinal Survey of Women and Families (KLoWF)**

*Covariates*. Number of siblings is measured as a continuous count of total siblings reported, capturing family size and potential resource dilution in childhood. Birthplace urbanicity classifies where the respondent was born into four categories: large metropolitan area; mid-sized or small city; rural area; or overseas. This serves as a proxy for early geographic context, including access to schools, labor markets, and health care. Mother's education at around age 15 is coded as a binary indicator for whether the respondent's mother had completed high school or more versus less than high school. Family financial situation at around age 15 is based on a self-reported assessment of how the respondent's household was doing financially in adolescence, with higher values indicating better perceived household economic conditions. Parental relationship quality at around age 15 is also a self-reported measure of how well the respondent's parents were getting along with each other when the respondent was a teenager, with higher values indicating better relationship quality.